\documentclass[preprint2]{aastex}

\slugcomment{Not to appear in Nonlearned J., 45.}

\shorttitle{Radio galaxy morphologies}
\shortauthors{Saripalli}

\begin{document}

\title{Understanding the Fanaroff-Riley radio galaxy classification}

\author{L. Saripalli}
\affil{Raman Research Institute, Bangalore, India}
\email{lsaripal@rri.res.in}

\begin{abstract}
The simple, yet profoundly far-reaching classification scheme based on extended radio
morphologies of radio galaxies, the Fanaroff-Riley classification has been a cornerstone
in our understanding of radio galaxies. Over the decades since the recognition that there are
two basic types of radio galaxy morphologies there have been several findings in different 
wavebands that have reported properties on different scales. 
Although it is realized that there may be intrinsic as well external causes an overarching view of
how we may understand the two morphological types is missing.
With the radio power-absolute magnitude relation (the Owen-Ledlow diagram) as backdrop we review and 
develop an understanding of the two radio galaxy types in the light of what is known about them. 
We have for the first time included the dust properties of the two FR classes together with the 
relative orientations of dust, host major axis and the radio axis to present a qualitative framework within
which to understand the conditions under which they form. We discuss how the host elliptical
and its history can explain the distribution of radio galaxies in the Owen-Ledlow diagram.
Mass of the host elliptical galaxy is a crucial player in deciding what type of a radio galaxy it can host
in what conditions. Benign conditions, characterized by natural evolutionary processes, 
most easily give rise to FR-I type sources in ellipticals 
of all mass regimes where as with FR-IIs we reason that it is hard to form them without mergers. 
In undisturbed conditions elliptical galaxies appear to acquire stable states where the black hole axis settles
along the host minor axis. With the steady conditions and 
the continuous supply of ambient gas, the FR-Is in principle, 
may be powered for a long time. 
Aided by mergers and interactions low mass ellipticals more easily form FR-II type sources than more
massive hosts.
LEG FR-IIs may include dying as well as restarted FR-II radio galaxies.

\end{abstract}

\keywords{galaxies: elliptical and lenticular, cD --- galaxies: active --- radio continuum: general 
--- galaxies: jets --- galaxies: nuclei}

\section{Introduction}

Although radio galaxies come in a variety of morphologies a basic division in 
structural types has stood the test of time: the Fanaroff and Riley classification
\citep{fan74}. It was based on the recognition that radio sources often came in two 
flavours, those having edge-darkened (FR-I) morphologies and those having edge-brightened (FR-II)
morphologies. The division was based entirely on radio structures that exist on large
scales from few to several hundred kiloparsec to several megaparsec. That the two 
radio morphologies also divide on the basis of several other
properties both on large scale and small points to more fundamental processes at work
that can have profound effect on what type of structures form on large scales. 

Since the
time of this simple classification scheme there have been several
attempts to understand how these two basic morphologies arise. Building on the basic
models of \citet{bla74}, \citet{sch74}, and \citet{fal91} there are detailed models today that are able to reproduce with
fair consistency data gathered on these source populations 
\citep{kai97, blu99}. At the heart of any model that seeks to reproduce the large scale radio structures 
is the basic tenet of interaction between a jet and its ambient medium. The interplay between the
two under varying characteristics of the jet and the external medium through which it propagates 
is shown to produce the two types of structures. Although other 
source morphological types have been discovered they have usually been understood within the
Fanaroff-Riley classification. 

While the two types of radio structures are understood as arising out of the different interactions 
that low and high power jets have with the environment \citep{sch74, bic86} a basic
question remained however as to whether the two source types were intrinsically the same or different: 
could the FR-Is and FR-IIs be the consequence as well of very different central engines and host galaxies 
or the consequence entirely of the different interactions with the environments? 

Over the several decades a number of multi-wavelength studies on FR-I and FR-II radio sources
has gathered a variety of data on these sources testifying to the importance of the question lying at
the focus of these efforts. Several important findings related to the two categories of sources
were reported. A picture is emerging which highlights the
intimate connection between the elliptical galaxy and the radio galaxy it hosts, where there is a 
dynamic relationship between them and where the history of the host galaxy has an important role. 

In this paper we develop a framework that brings together some key research in this area
that was not considered earlier. There have been efforts in the past, particularly the work of
\citet{bau95} who have put forward a scenario for forming the two types of radio galaxies. We make use
of their research as well as more recent findings in our effort to present a coherent picture in which 
we may understand the conditions in which the two basic morphological types form. 
We summarize some of the key observations of FR-Is and FR-IIs in section 2 after which in sections 3---6 
we introduce and examine properties of FR-Is and FR-IIs not considered earlier, which we now incorporate within 
our framework.  In section~7 we present and 
discuss the framework where due attention is
given to the insightful Owen-Ledlow diagram \citep{owe93, owe94} and the crucial role it has in the effort to
understand radio galaxies. This is followed by the summary.

\section{A summary of some key findings on FR-Is and FR-II radio galaxies}

In this section we list some of the well-known and well-used findings reported for
the two classes of radio galaxies. 
Right in the seminal paper recognizing the two radio source morphologies, \citet{fan74} 
pointed out the correlation of the morphologies with the total radio power:
the two morphologies are divided in their
total radio power with the edge-darkened FR-I type radio galaxies having lower radio powers and the 
edge-brightened FR-II type radio galaxies having higher radio powers. 
The two radio source types were found to be divided at the so-called
dividing power of $10^{25}~W~Hz^{-1}$ at 1.4~GHz. 

FR-Is and FR-IIs also compare interestingly with respect to their environments. At nearby redshifts ($z < 0.5$) 
while the FR-Is are found to be located in 
dense environments (galaxy clusters) the FR-IIs are often found to be hosted by field galaxies.
At higher redshifts though both FR-I and FR-II type radio sources are found in rich environments
\citep{hil91}. The elliptical galaxies that host the two FR types were also found to show differences.
Broad band imaging of the host galaxies showed that hosts of FR-IIs were found to be bluer than hosts of FR-Is and
often showed signatures of mergers \citep{hec86, smi89, bal08, ram12}.
Then again the host galaxies of FR-Is were found to be more massive than the FR-II hosts
\citep{owe89, gov00}. While AGN optical spectra of some of the FR-II hosts showed emission lines 
this was almost never the case in FR-I optical spectra. We will return to this topic of AGN spectra
later in this section.

One of the important developments in this area has been the Owen-Ledlow diagram \citep{owe93, owe94}
where the two source types divide about a line with slope of about 1.8 in the 
total radio power-absolute optical magnitude plane. 
The dividing power, hiterto
believed to be fixed was instead found to be increasing with the host optical luminosity. 
If FR-I structures are a result of jets that are affected by entrainment, instabilities and turbulence
then one could understand the increasing dividing power with absolute host optical magnitude
as galaxies with jets that found it increasingly difficult to remain collimated and supersonic as the ambient
medium became denser \citep{bic95}. This dependence of the FR-I/II dividing power on the host
absolute magnitude put the spotlight on the role of environment in producing the two 
different radio source morphologies. 

The finding of a hybrid morphology in some radio galaxies, with one lobe being edge brightened and 
another edge darkened \citep{gop00} strengthened the view that environment in which the jets propagated 
were ultimetely responsible for the kind of structure that developed on large scales. 

With the increasing availability of high resolution optical imaging as well as more complete optical
spectroscopic observations, details with respect to the AGN as well as host galaxy characteristics on 
more global scales emerged that had to be understood. 

The study noting differences in optical spectral 
properties of FR-I and FR-II hosts reported by \citep{hin79} was followed by works that used
classifications based on spectral line ratios \citep{bau92, lai94, tad98, chi02, bot10} which clearly showed that the
two morphological types exhibited different behaviour. While FR-I radio galaxy hosts always exhibit
optical spectra with only absorption lines or O[III]/H$\alpha < 0.2$ (low ionization emission line radio galaxies, 
LEGs, following the definition of Laing et al.), the FR-II hosts were of mixed 
category. Some FR-II hosts were like FR-Is with either only absorption lines or prevalence of low ionization emission 
lines with low O[III]/H$\alpha$ ratios but some others showed spectra with strong high ionization emission lines 
with O[III]/H$\alpha > 0.2$ (high ionization emission line radio galaxies, HEGs). 

\citet{bau92} found clear differences in the 
characteristics of the emission line gas in the two FR types. Most FR-IIs appeared to have rotating 
disks of line-emitting gas on large scales up to 15 kpc that also sometimes included disks with chaotic
and turbulent motions which contrasts with those in FR-I type sources. These findings were linked with 
two different modes of gas acquisition for fueling the AGN in the two FR types. 
The quite different emission line characteristics of FR-I and FR-II radio galaxies led \citet{bau95} to
explore different ways of producing the two morphologies. A preferred model was that the differences arose in the
different accretion rates: low accretion rates in FR-Is and high accretion rates in FR-IIs. Differences
in the black hole spin too was suggested for the two classes with FR-Is having a lower black hole spin.
\citet{mar04} also found that the accretion
rates needed in FR-Is were very low, less than $\sim 0.001$ in Eddington units. However again interestingly
the FR-IIs appeared to span two regions, one that had similarly low accretion rates as FR-Is and a small
fraction that required higher accretion rates. 

X-ray and IR observations (\citet{har07} and references therein and \citet{bes12}) and the
optical spectroscopic studies \citep{bot10, mah11} further support and highlight 
the division based on line ratios (LEG and HEG type) and the link with accretion mode and source of fuel. 
Increasingly it is being suggested that the two optical spectral classes are powered by 
two different accretion modes and fuel sources: radiatively inefficient low accretion rates sourced from 
hot gas accretion that powers
the low and high power LEGs where as radiatively efficient high accretion rates sourced from cold gas accretion 
powering the HEG sources. The hot gas is suggested as originating from 
the hot coronae of the hosts and from the stellar mass loss from the stars in the galaxy. On 
the other hand the cold gas is suggested as originating from a gas-rich merger with another galaxy.

High resolution optical imaging with HST 
revealed the prevalence of nuclear optical cores in both FR-Is and FR-IIs \citep{chi99, chi00, chi02}. 
However there were clear differences between the two FR types, with a clear correlation seen 
between the optical core powers and the radio core powers for the FR-Is and a more complex behaviour
for the FR-IIs. While the FR-1s were inferred to have unobscured, radiatively inefficient accretion disks
the FR-IIs once again divided into two distinct types: a population that showed similar properties as the
FR-Is (constituted by the LEG FR-IIs) and a population constituted by the HEG FR-IIs where presence of
dust torii and radiatively efficient accretion disks were inferred. 

In the following sections we look at some additional properties of the two FR types where they show differences
and we try to develop a framework inclusive of these findings. We first consider the dust characteristics 
of the hosts of the two FR types.

\section{Dust and FR classification}


More than three decades ago \citet{kot79} reported the perpendicularity of dust
distribution and radio axes for a small sample of radio galaxies. At that time the
sources involved only the nearby galaxies with prominent dust distributions. It was
also the case that the radio sources were mostly FR-I radio galaxies. With the HST
making possible the high resolution imaging of galaxies, dust on much smaller scales 
was revealed and in particular dust distributed in the nuclear regions. 

\citet{van95} and \citet{ver05} used archival HST data to study dust in elliptical galaxies. They reported a 
high prevalence of dust among the hosts of radio galaxies as compared to normal ellipticals that
suggested a link between dust and nuclear activity. Differences in properties were seen for dust on small ($<250~pc$)
and large scales ($>250~pc$). The only dust features to show kinematic coupling with the stars were those on 
small scales (4 out of 7; Fig.~7; \citet{van95}) where as large scale dust showed
no such kinematic coupling. Large
scale dust (and at least some small-scale dust) therefore carry signatures of angular momentum
different from that of the host ellipticals and indicate external origin, perhaps in mergers. 
Mostly consistent with the differences are the properties of appearance and location of the dust on large and small 
scales. The dust on large scales generally had an irregular appearance and was distributed uniformly with respect to 
the major axis where as the dust on smaller scales was mostly of regular, relaxed appearance and was
located on the projected major axis of the host galaxy \citep{van95, ver05}. 

\citet{dek00} and \citet{deR02} used HST observations of two well-known
and well-imaged (both in radio and the optical) radio galaxy samples (the 3CRR and 
B2 samples respectively) to study the dust characteristics of the radio galaxy host ellipticals. 
These observations revealed the very different dust characteristics of the two FR types.

While both de Koff et al as well as de Ruiter et al reported the trend for the dust 
to be distributed perpendicular to the radio axis it was found to be the case predominantly for the FR-I 
radio sources. The FR-IIs on the other hand showed less or no tendency for perpendicularity. The dust 
in FR-Is was mostly seen in the form of small regular circumnuclear dust lanes or disks where as the
dust in FR-IIs had varied morphology and extent. 
Both also reported 
the lower dust masses among FR-Is as compared to FR-IIs. Interestingly, \citet{deR02} noted that the 
perpendicularity between the radio axis and dust is confined to lower
power FR-Is, becoming weak or absent in stronger FR-Is. \citet{dek00} noted that in FR-IIs dust-radio
perpendicularity was seen only when dust was 'concentrated close to the nucleus'. 

While the dust-radio axis relation and dust characteristics 
such as appearance and size have been examined in all the above works, the location of the dust (perpendicular
to the radio axis) with respect to the host major axis got little attention except in the work on normal and 
FR-I ellipticals by \citet{ver05}. Their Fig.~4 shows that in several cases  where the dust is perpendicular to
the radio axis it lies within 25 degrees of the host major axis (8 out of 14 sources). The coincidence with the 
major axis is even more impressive when only smooth, regular dust ellipses are considered (6 out of 7 sources). 
The persistence of dust-radio perpendicularity even when dust does not lie on the host major axis, as seen in some 
of the cases, is also reported by \citet{van95} (in Fig.~8) where although the dust is 
perpendicular to the radio axis in 6 galaxies (admittedly, a rather small sample) it lies on the host major axis in 
only half the sources. 

We point out that the detection as well as appearance of the dust features will depend on distance to the source. 
Since FR-Is and FR-IIs are generally selected from different redshift regimes (FR-Is near and FR-IIs far) 
could the differences in dust characteristics of the FR-types be attributed to distance related effects?

The dust characteristics of radio galaxies being discussed here have been sourced from mainly three previous 
works, \citet{dek00}, \citet{deR02} and \citet{ver05}. All three authors have 
endeavored to study the effect of distance on their findings. Little effects of distance are found for 
radio power-dust mass relation or dust morphologies, classification, or position angles.

To collect together the points relevant for the dust-radio axis perpendicularity relationship, it appears that
the relation is strongest for lower power FR-Is with jets rather than for FR-IIs and that dust need not always
lie on the host major axis. 

As for the dust in elliptical galaxies, without concerning ourselves with whether it hosts
a radio galaxy or not, it seems that dust can exist on large scales and small scales and the morphologies of the dust
on the two scales is different. The dust on large scales is mostly unsettled where as dust on small scales appears
mostly regular and settled. The division of large-scale dust and small-scale dust seems to largely adhere to
the division of the FR type with FR-II radio galaxies associated with dust on large scales
and FR-I radio galaxies with dust on small scales.

Dust however is clearly found to be important for the AGN activity and the dust mass is clearly found to be 
related to the radio power with increasing dust mass for higher power radio galaxies going from low power
FR-Is, high power FR-Is and to the FR-IIs.

Table~1 puts together the dust properties of FR-I and FR-II sources.

\begin{deluxetable}{ll}
\tabletypesize{\scriptsize}
\tablecaption{Summary of dust properties of FR-I and FR-II sources}
\tablewidth{0pt}
\tablehead{
\colhead{FR-Is} & \colhead{FR-IIs} 
}
\startdata

Distributed on small scales ($<250 pc$) & distributed on large scales ($>250 pc$)\\
Circum-nuclear & not circum-nuclear\\
Sharp, disk-like & irregular and filamentary\\
Mostly on the host major axis & no relation with the host major axis\\
Small dust masses & large dust masses\\
Perpenducular to radio axis & no relation with radio axis\\

\enddata
\end{deluxetable}

Therefore there is a clear division seen among the dust characteristics of the two FR types. Any framework
for understanding the FR categories needs to consider these rather stark differences. The dust properties appear
linked to the kind of radio morphology that emerges on tens to hundreds of kpc scales. 

If dust properties on global scales are found to be related to the manner of manifestation of the twin
jets from the central engine, whether an edge-darkened or edge-brightened extended radio source
results, it is worth examining if dust origin can be linked to the formation of the two source types.

\section{Relative orientations of radio and host optical axes and the FR classification}

\citet{sar09} studied the relative orientations of radio and host optical axes of the 
3CRR FR-II sample and found that there was no preferred relation; the radio axes are distributed over a wide range of
angles from 0 to $90\arcdeg$ with respect to the major axes. Several previous studies also reported the same lack of 
relation between the radio and host optical axes for FR-IIs \citep{pal79,gut80,san87}. Curiously, it was only more recently
that this study was performed for FR-I type radio galaxies. \citet{bro11} used
a large sample of FIRST radio sources identified with SDSS galaxies for which they derived the radio and
host optical axes orientations and they reported a strong tendency for radio axes in their (largely FR-I radio galaxies)
to be oriented along the host minor axis. Interestingly, this commonality of radio and host minor axes 
position angles is not shown by the stronger of their FR-I sources. 

Once again these clear differences between the relative orientations of radio and host optical axes need to be
brought within any wider framework seeking to understand the two FR types. 

\section{The host galaxies of FR-I and FR-II radio galaxies}

Elliptical galaxies have been the subject of numerous studies. Given that they form the hosts of 
radio galaxies it is only natural to examine some of the recent detailed observations of this family of
galaxies for any bearing they may have on the formation of the two radio galaxy types. While we do not 
attempt to summarize the body of data gathered on elliptical galaxies we use some of the more recent
detailed observational studies to highlight aspects that clarify the characteristics of radio galaxy hosts.
It is now well established via a number of works that elliptical galaxies fall into two categories: the
fast rotators and the slow rotators (\citet{ems07}, \citet{kor09} and references within). These two classes of ellipticals 
have distinct properties in several respects. Of relevance to the hosts of radio galaxies we note that the
dividing optical luminosity for the fast and slow rotators is at $M_{b}=-20.5$ with slow rotators being more
luminous. With the hosts of radio galaxies predominantly being as bright or brighter than $M_{b}=-20.5$ it appears 
that most of the radio galaxy hosts fall in the group of slow rotators. This is supported by two other 
properties of radio galaxy hosts: their ellipticities and masses. Slow rotators are found to have
ellipticities flatter than 0.3 which is also the range estimated for a large fraction of 3CRR host
ellipticals \citep{sar09}. Moreover the slow rotators are also the more massive with masses larger
than $10^{11} M_{\odot}$ which is also the mass regime for radio galaxy hosts. 

There are several other distinctive characteristics of the two groups of ellipticals but we confine ourselves
to the recognition that the hosts of radio galaxies resemble to a fair extent the slow rotating, massive
elliptical galaxies. While this galaxy group is not as strongly oblate as the fast rotators the 
difference in the position angle of the photometric and kinematic axes is mostly within $20\arcdeg$ for the
sample studied by \citet{ems07} with as many as half having values within $10\arcdeg$ (their Figure~4). 
It is reasonable to expect that a large fraction may be close to being oblate although there are clearly known 
triaxial galaxies.  

\section{In Perspective}

We now attempt to put the dust properties of FR-I and FR-II sources in perspective. First, the scale lengths
that we are dealing with regarding the dust on one hand and the location of the jet origin and hence the
black hole spin axis are entirely different. As noted by \citet{van95} one does not expect there to be any 
correspondence between the jet direction and the inflowing material considering that in the vicinity of
the black hole frame dragging precesses the orbit of the incoming fuel. Even with the resolution of
the HST observations the seen dust is on scales nearly 7 orders of magnitude larger than the accretion
disk scale. Yet it is with this large scale dust disk that a close relationship exists between
it and the jet axis and that too predominantly only in FR-Is. 
Now, dust as we have seen is important for
AGN activity. Dust exists also in FR-IIs and in fact dust masses in FR-IIs are found to be larger than in
FR-Is. Although, as seen in section~5, the host ellipticals of FR-IIs and FR-Is are similar, having similar range of
absolute magnitudes and ellipticities yet the dust in FR-Is can appear different and also behave differently
in its relationship with the radio axis: we need to understand why FR-Is are being singled out for the 
$\it{dust-radio~perpendicularity}$ relation and why the dust appears different in FR-Is and FR-IIs. The
reasons as to why the dust-radio axis perpendicularity exists were speculated upon by \citet{ver05}. In 
attempting to understand the reason for the particular relationship preferentially shown by the FR-I 
radio galaxies and in particular the lower power FR-Is we will also need to understand why the
correlation is weaker or not seen among the FR-IIs. We will address this in Section~7.

Then again, with regard to the radio-host major axis perpendicularity relation it is the FR-Is that show this
and not the FR-IIs. If the host galaxies of the two FR types are similar sharing the same range of 
optical luminosities and ellipticities we need to understand why FR-Is are being singled out for the
$\it{radio-major~axis~perpendicularity}$.

We may express in other words to say that predominantly in FR-Is there is some kind of equilibrium 
configuration set up between the host galaxy, the dust and the black hole at the centre. Whereas although
hosted by ellipticals of similar type such an 'equilibrium' evades the FR-IIs. It appears that more
powerful the beams less are the equilibrium relations with the dust axes and host major axes respected.

Below we will attempt to understand the conditions that may be causing these differences between FR-Is and
FR-IIs.

\section{Understanding the two source types}

\subsection{The Owen-Ledlow diagram}
\citet{bic95} developed the theoretical basis for the dependence of FR-I/II dividing power on the absolute
magnitude of the host elliptical (the Owen-Ledlow diagram). The FR-I/II borderline jet energy flux (and hence 
total radio power) was related to the host optical magnitude through parameters that affected the jet 
propagation i.e. the central pressure. A given
absolute optical magnitude of an elliptical galaxy (or a given mass of elliptical galaxy) has an 
ambient pressure which corresponds to a unique transition jet flux such that for higher ambient pressures 
(and hence for more luminous elliptical galaxies) the FR-I/II jet transition occurs at higher jet energy flux.
There have been other attempts to relate the jet advance with the ambient medium within the host galaxy 
(for example, \citet{kaw09}) and also attempts to understand the Owen-Ledlow diagram by redrawing the relation 
in terms of intrinsic parameters: nuclear photoionizing luminosity versus the black hole mass \citep{ghi01,wol07}. 

A point to note about the Owen-Ledlow diagram is the scatter along the Y-axis. One sees that no longer 
is it that a galaxy of a given optical magnitude and hence mass produces a radio galaxy of a fixed
morphology or fixed power. It is exciting to note that
galaxies of the same absolute magnitude can host FR-Is as well as FR-IIs and also FR-Is and FR-IIs 
having a range of different powers. It follows that under 
different conditions an elliptical galaxy can host an FR-I or FR-II or an FR-I of a different power and an
FR-II of a different power. 

The Owen-Ledlow diagram is a powerful representation of the story of radio galaxies, of the 
relation between the large scale radio structures and the galaxies that host them, revealing that for
every host galaxy mass there is a threshold jet power which is needed to produce an edge-brightened 
structure and that if conditions within change the same galaxy can host a radio galaxy of a different 
power or different morphology in its lifetime. The Owen-Ledlow diagram therefore incorporates within it the 
possibility of restarting of nuclear activity. We will see below that the simple diagram can also provide
a platform for understanding various other findings related to the two morphological types.

The translation of FR classes along the Y-axis was already remarked on by \citet{led96} and \citet{led97} 
who pointed out that this could be suggesting possibility of a transition between the two populations. 
In their preferred scenario for understanding the two FR classes \citet{bau95} also pointed out that 
the differences in the black hole spins (with low spin for FR-Is and high spin for FR-IIs) allowed for the
possibility of transition from FR-II to FR-I type. Such a transition is not unreasonable to expect since
radio galaxies have finite life times and the beam switch-off process occurs more likely over a drawn out
period of time rather than abruptly (as also discussed in \citet{sar12}). 
\citet{sar09} invoked such an FR-II to FR-I transition in morphology
to explain the five, restarted X-shaped sources that on one hand showed main-lobe, edge-darkened structures while 
exhibiting properties similar to the rest of the X-shaped source population all of which were of FR-II type. Indeed 
in using Monte-Carlo simulations to test whether observed samples of radio galaxies can be random selections of 
elliptical galaxies, \citet{sca01} found that the two FR classes are hosted by ellipticals extracted
from the same population. Also, the Owen-Ledlow diagrams of radio samples in later studies are not found to be 
as sharply divided between the two classes as originally found and there is non-insignificant number 
of FR-IIs below the canonical dividing line \citep{led96, lin10}. These below-line FR-IIs, 
that are also found to be compact in physical size, have been speculated as being FR-IIs that are likely to evolve 
into FR-Is as their low power jets get frustrated in interactions with the ISM of the host galaxy \citep{kai07}.
While these compact FR-IIs may be low power and young radio galaxies there may also be, as found among the
ATLBS-ESS sources \citep{sar12}, larger FR-II sources that are either relic-type or even restarted FR-II sources 
(where the outer lobes have reduced in power; Thorat et al, in preparation). The existence of FR-I type sources, the 
intriguing FR-I quasars, lying above the dividing line has also been reported \citep{hey07}. We discuss this
in later in this section.

The growing impression therefore is of flexible physical conditions across both the absolute magnitude axis as 
well as the integrated (radio) power axis. FR-I/II radio galaxies can both be hosted by elliptical galaxies having a wide 
range in absolute magnitudes (or masses) within the massive galaxy regime and at the same time not only do both 
FR types occur in galaxies of the same magnitude but FR-II radio galaxies can have integrated powers that lie 
below the dividing line. While the transition jet energy
flux could be unique to an elliptical galaxy of a given absolute magnitude it remains to 
identify the physical conditions as well as connection with galaxy mass which will determine the
type of extended radio morphology that will result.  

\subsection{Host galaxy mass and the FR type}
With the backdrop of the Owen-Ledlow diagram in mind we sketch the following scenario to understand the
differences in the properties of FR-Is and FR-IIs. Massive elliptical galaxies 
will have enough stellar mass loss to sustain a radio galaxy \citep{dim03,ho09}. 
Massive ellipticals will therefore frequently host radio sources given the regular source of fuel.
However because of 
the higher central gas pressures and more extended interstellar medium of more massive elliptical
galaxies, most jet powers generated will find it hard to retain their thrust over long distances
through the ISM \citep{bic95}. Only the more powerful of jets will remain supersonic and form FR-II morphologies. 
The resulting morphologies will therefore tend to be dominated by FR-I rather than FR-II morphologies. Also
high power jets (capable of negotiating successfully the increased ISM) 
may likely be less often produced than low power jets since such jets would require higher accretion rates 
than that generated 'in-house' (otherwise most ellipticals will
be associated with FR-II sources) and in the case of massive ellipticals will likely need gas-rich mergers.

The inability of massive ellipticals to host FR-II morphologies gets more and more aggravated with
increased mass of the galaxies. This and the high frequency of association of massive ellipticals with radio sources 
is consistent  with the finding by \citet{bes05} that the 
fraction of low power radio sources hosted by elliptical galaxies is high (as high as $30\%$) and 
this fraction increases with galaxy mass.

It follows that lower the mass of the host galaxy it gets easier to host radio galaxies with
FR-II morphologies. Unlike the more massive ellipticals the ISM is less rich and less extended and the threshold 
jet power is lower. However, while there is still the stellar mass loss that
is available as fuel, it is likely to be available at lower rates than in the more massive ellipticals
resulting in only weak sources. 
Overcoming these lower threshold jet powers may not need 
a large jump in accretion rate however and they may relatively easily be breached by even small ingestions of external 
fuel.
A dependable way for a 
lower mass elliptical to create radio galaxies having FR-II morphology is if this additional source of 
fuel comes in, say, through a merger.

Host galaxy mass and its history appear to be key to the type of extended morphologies that result on large scales. 
Since the 
galaxy mass correlates with the central black hole mass how do black hole mass differences bear on the two radio 
galaxy types? Black hole mass has a direct consequence to the Eddington accretion rate. With in-house accreted 
mass also scaling with the galaxy mass differences in the Eddington ratios could arise more from causes such 
as e.g. merger opportunities, active star-formation or environments.

\subsubsection{Understanding the dust characteristics of FR-Is and FR-IIs}

The reasoning given above accounts for several of the properties that are found for FR-I and FR-II
radio galaxies. For example, with massive and mostly oblate type ellipticals more likely to be hosting 
FR-I radio morphologies it explains
the large host masses reported for FR-Is. With stellar mass loss and  the
gas from the hot coronae being the predominant fuel source \citep{bot10}
one expects that there is a regular supply of gas and dust. This dust shares the angular momentum of the 
stars in the galaxy and unless there is a merger it will make its way to the centre of the deep potential well
unmolested. With the major axis plane being the equilibrium plane in oblate-type ellipticals the dust will
tend to settle there where it can form stable closed orbits \citep{van95}.
The rate of supply of gas and dust is steady and is only as high as the rate at which the 
stars evolve or the coronal gas is ingested. This decides the upper limit to the disk accretion rate. The
relatively low rates ensure low power jets which, given the high galaxy mass, will result in FR-I radio structures. 
Given the stable nature of the process of accumulation of gas and dust it will settle into
a regular disk at the centre of the galaxy. Moreover the dust will get to be observed only when it has accumulated
in sufficient amount which happens when it reaches smaller scales near the central regions. The internally 
originating dust is expected to be generated from regions uniformly distributed over the galaxy without
having visible signatures such as clumps or disks or lanes
and hence it will not be observed. Hence the correlation of dust morphology with location on small scales and
location on the major axis. In such a picture, FR-Is are predominantly seen associated with massive ellipticals and 
without the need for mergers.  We point out that this scale on which the dust is seen is still several orders of
magnitude larger than the accretion disk in the vicinity of the black hole. In such a 
steady state within the galaxy where dust has been collecting at the centre for a long period the black hole 
would have been re-aligned perpendicular to the dust disk. 

As for the FR-II sources, their hosts are more likely to be small and to have swathes of dust of irregular and 
filamentary structure given the likely merger history that we reasoned was needed to preferentially produce an FR-II. 
The unsettled dust in FR-IIs is more likely to have been recently acquired given that dust settling time is estimated 
at $10^{8}yr$ \citep{van95}.

\citet{nat98} derived black hole re-alignment timescale for a case where an accreting black hole 
experiences a reverse torque due to the outer disk of the host galaxy. The re-alignment timescale 
is derived to be few $10^{5}~yr$ for an AGN radiating at a luminosity which is a tenth of the 
Eddington luminosity. If we use a more realisitic estimate of the luminosity applicable to 
FR-I radio galaxies (0.001 or lower; \citet{mar04, ho09}) the black hole will realign over a time scale
that is several orders of magnitude larger. Correspondingly the more powerful radio galaxies with their higher 
luminosities may tend to realign earlier (although still on timescales few orders of magnitude larger than
$10^{5}~yr$). The observation that FR-Is have radio axes often aligned with
the minor axes of their hosts suggests that they have been left relatively unperturbed for a long
enough time to have their black holes realign with the minor axes. However, for the FR-IIs, such undisturbed
conditions may have eluded them and the black hole may have been prevented from realigning with the minor axis. 

Dust-radio relationship appears to be stronger than dust-major axis relationship
in FR-Is. In several FR-Is the radio axis continues to be orthogonal to the dust even when the dust is 
not located on the host major axis (see Section~3). We have already noted that FR-Is are more than adequately sustained 
with the fuel which is in sufficient supply in massive galaxies.
The steady supply means that dust has had time to settle at the centre and form a stable
disk on the major axis. The dust-radio relation gets established in these calm conditions.
With dust-radio relation holding even when dust is not on the major axis (as seen in some FR-Is) it implies that any
dust that comes in from outside and is at sufficiently close distance to the central region 
has a strong effect on the black hole angular momentum and can re-orient it. 

At this stage we note that there are some remarkable similarities between the picture we are arriving at in
understanding the 
dust properties of the two FR types and the scenario sketched by \citet{bau92} to explain the properties of
the extended emission line regions. Already articulated clearly as "gradual, steady" feeding
in the case of FR-Is and "impulsive" feeding in the case of FR-IIs as inferred from the distinct extended 
emission line gas characteristics of the two classes, such a physical picture seems to also emerge from the dust 
properties. The association of large-extent emission line regions with their large rotational velocities (larger than
the stellar rotational velocities) and the large kinematic excursions with FR-IIs (whose hosts also exhibited
morphological distortions) strongly suggested that the gas may have been acquired in recent mergers. In contrast the
much smaller extents as well as lack of significant large scale motions in the emission line gas in galaxies that 
were almost all at cluster centres and hosting FR-Is suggested a local origin for the gas.

A potential upset for the model we have been developing for the FR-Is and FR-IIs is that the emission line gas
rotation axis and the radio axis in FR-IIs align within about $30\arcdeg$ \citep{bau92}. However as the authors
themselves describe, the emission line gas nebulae in the FR-IIs, "show asymmetric rotation curves, large velocity
excursions $\pm100~km s^{-1}$ from simple rotation, offsets of the kinematic and optical centers, broad lines, and 
mislignments 
of the rotation axis and the minor axis of gas distribution". It is indeed difficult to understand how in the
midst of a merger and with the gas, "not yet settled into an equilibrium orbit or the activity generated in the galaxy
nucleus has disrupted the orderly rotation of the gas" there is the tendency for alignment of the gas rotation 
axis and radio axis particualrly when little alignment with the dust and radio axis is observed for FR-IIs (with
dust and gas assumed to be present together). 

The details of the process of fuel accumulation are not clear. At what stage in the fuel accumulation at the
centre the activity is triggered is also not clear. However as low power FR-Is may need only a small amount of fuel 
to trigger them, they can also form in the 
early stages of ingestion of the externally generated fuel before the full accretion rate is reached 
or when generated in massive hosts (where there is a 
continuous supply of fuel which would have settled on the major axis) or in the late stages of an FR-II 
when the fuel is depleted but enough time has passed and the last vestiges of the dust has settled into a 
regular disk on the major axis. That the latter possibility cannot be the only way to form FR-Is  
was already ruled out by \citet{dek00}
since there is no difference in the dust masses estimated for large FR-IIs and small size FR-IIs, however it is
still a possibility in some cases.

With FR-IIs being preferentially hosted by smaller ellipticals aided by mergers or larger ellipticals also
aided by mergers (in the former because internally generated fuel is insufficient and then in the latter
because the internal resistance requires higher accretion rates) and with lower mass ellipticals being more
numerous given the Schechter luminosity function, we have a situation where in samples of FR-IIs the dust 
will be seen preferentially on larger scales in a filamentary form rather than as disks at the centre. 

What about the general lack of dust-radio perpendicularity in FR-IIs and the lack of relation between the radio axis 
and the host major axis in FR-IIs? With mergers aiding
the hosts in producing jets powerful enough to overcome the ISM and produce FR-II structures this lack of 
perpendicularity relations may occur if the black hole spin is affected by the merger,
whether by the triggered gas inflows or black hole-black hole merger \citep{hop11}. The likley long re-alignment 
timescales would ensure that the black hole axes (and hence the radio axes) show no relation with the incoming dust.
We note that at the AGN scales there 
should be the expected perpendicularity between the accretion disk and radio jets but observations only 
pick out the larger-scale dust. The fueling is more dynamic in FR-IIs
where the externally originating dust and gas could come in tranches of possibly differing orientations between them 
as also with respect to the major axis. The large-scale dust picked up in observations will therefore be
of filamentary morphology and distributed with little correspondence with the radio axis. 
The same can be the case with high power FR-Is. These would be more likely generated in 
massive ellipticals that have had a merger that brings in more fuel than what the internally generated material
brings. 

At this stage we bring attention to the fact that in the several examples of restarted FR-II radio galaxies
a change in axis between the two activity epochs is rarely observed. This can be taken to infer that the 
black hole spin axis remains steady between the two epochs. How can we understand this in the light of the
reasoning for the observed lack of correlation of radio axes with either the dust or the host major axes
in FR-IIs? It is clear that while mergers could have perturbed the black hole axes in FR-II hosts whatever 
is responsible for the interruption and re-triggering of the AGN, it has been gentler.
It is possible for the black hole axis to remain steady within a merger event if we can 
associate the interruption and restarting of activity with the interruption to the fueling as each tranche
of fuel is exhausted. Admittedly timescales are important here, for example the timescale over which the
dust (gas) segment gets depleted and the re-alignment timescale. In the context of timescales we point out that 
(based on the lack of relation between the dust and radio axes in FR-IIs) the AGN may be triggered even as the 
much of the dust is settling and since there is no trailing radio emission seen, the 
entire AGN beam activity (including the quiescent phase and renewed activity) 
may be happening on a timescale small compared to the realignment timescale.

The "gradual, steady" feeding of the central engine in FR-Is with potentially limitless supply of fuel 
suggests that the fueling can remain active for a long time, perhaps much longer time than in FR-IIs. 
In these conditions it remains to understand how there can be any interruption to the activity in FR-Is.

There will be cases where the jets are oriented along the host minor axes. With less ISM to propagate through 
the powerful jets will advance more easily. These minor axis
sources - as opposed to the major axis sources - will have a tendency to host a larger fraction of large size 
radio galaxies. In our earlier work \citep{sar09} we reported such a tendency in the 3CR 
sample and we also reported the tendency for giant radio galaxies to have radio axes oriented along the host 
minor axes. 

The richer environments of FR-Is may be environments that are conducive to creating edge-darkened
structures because of the higher resistance that they offer the jets as they propagate out. 
However for FR-IIs, as we have already reasoned, a reliable means of generating them is via the lower mass ellipticals 
that have undergone mergers so low mass ellipticals 
that have had no merger history but reside in rich environments should have the least probability to host jets 
powerful enough to create FR-II structures. 
The more frequent mergers at higher redshifts means that there is a higher availability of cold gas and this is
conducive to the formation of FR-IIs. 

In the steady fuelling conditions inferred to be prevalent in most FR-I hosts the central engine may, in principle, 
never cease activity or at least may continue to remain active for a long time. However the multiple X-ray cavities
observed in several galaxy clusters reveal a different picture that implicates multiple episodes of AGN activity 
implying an unsteadiness of the AGN activity in these otherwise calm conditions. A missing factor is the 
feedback effect of the radio lobes which exercise control on cluster scales finally feeding back to the fueling
itself (\citet{mcn07} and references therein). With nearly every cooling flow cluster hosting multiple cavities the 
causes of AGN episodicity appear to be a feature of and linked to the specific conditions in a regulatory manner that
can itself be a periodic phenomenon subject perhaps to disturbances to the large scale steady flows operating
in the cluster. 

\subsubsection{The classification based on optical spectra}

The traditional classification of radio sources based on their large-scale morphology is increasingly being 
examined in light of the central AGN spectroscopic properties. The radio morphologies may be viewed as 
consequences of the central AGN properties influenced or mediated by the prevailing conditions and 
state of the AGN (whether active, or waning or dead) and the environment through which the jet propagates.


As for the optical emission line characteristics of FR-Is and FR-IIs  
if all high ionization emission
line radio galaxies are of FR-II morphologies, all FR-Is are low ionization emission line galaxies and some 
FR-IIs are low ionization emission line galaxies, then from the strong relation between the radio 
luminosity and the narrow-line luminosity it appears that the HEG FR-IIs are among the most powerful of the 
FR-II population (however also see \citet{chi02,bot10}).
We will infer then that there is a high threshold jet power above which HEG characteristics manifest. This 
threshold power is higher than the dividing power at any optical host luminosity (to exclude FR-Is). 

What then of the LEG FR-IIs? How do we understand a LEG FR-II? This was also the question 
raised and discussed by \citet{lai94} and \citet{chi02}. They gave several reasons to support the view that 
LEG FR-IIs are an isotropic population with at least a part being the parent population of BL-Lacs. 
With their close resemblance to FR-Is 
in their optical nuclear properties \citep{chi02} the FR-Is and LEG FR-IIs share a state of the AGN 
characterized by a weak central ionizing source (radiatively 
inefficient disk) and lack of any substantial gas or dust torus. It is tempting to view this class of FR-IIs as 
inclusive of sources that have waned in AGN activity or even FR-IIs 
that have restarted at only small accretion rates. The mixed characteristics of FR-IIs (HEG and LEG type) 
may be reflecting the variable central engine conditions. On the other hand the mixed radio morphologies observed in 
FR-Is has led to the 
speculation that at least some lobe-type FR-Is may be dying FR-II radio sources \citep{sar12}. 

In \citep{chi00} a small number of FR-IIs were reported clearly showing optical core properties
indistinguishable from FR-Is. On examining the radio structures of the five FR-IIs one is of FR-I morphology
and three have characteristics that classify them closer to being relic sources: lobes with very low axial ratio and
at least one of the lobes a relic lobe and weak hotspots. One other is a classic X-shaped source with bright hotspots
in its main lobes. These five sources (one is an FR-I) are also reported to be in cluster environments. 
It is likely that the three FR-IIs (2 are classified LEG-type and 1 HEG) with nuclear properties similar to 
FR-Is and radio morphologies that are more 
non-classic FR-II type are sources where the accretion rate has reduced to a level where it is not high enough to 
sustain an FR-II morphology and the radio morphology is showing the signatures of a dying FR-II. 

On examining the morphologies of the FR-II radio galaxies in \citet{bot10} 
although both classes of FR-IIs (the LEG and HEG FR-IIs) include a mix of source characteristics the LEG FR-IIs are 
associated with a larger fraction of sources with restarted AGN (3C293, 3C236, 3C388) or relaxed double
type (3C310, 3C401). In contrast the HEG FR-II mostly include sources with bright hotspot lobes and X-shaped
radio sources (3C136.1, 3C223.1, 3C403). The LEG FR-II group does not include any with X-shaped morphology. If X-shaped
structures are a result of deflection of backflows \citep{sar09} it is not surprising that these sources are predominantly 
found to be associated with HEG FR-IIs, sources that are characterized by bright hotspot lobes capable of generating
strong backflows.

It is not necessary for all LEG FR-IIs to also experience hot gas accretion for their sustenance 
like FR-Is. Given the different response times of the sub-galactic narrow line regions and the
several hundred kiloparsec-scale regions of synchrotron plasma it is possible that a current slowed-down state 
of the accretion can reveal an edge-brightened source on extended scales with an associated LEG AGN.

Besides having FR-I like AGN the LEG FR-IIs share more properties with the FR-Is: they
have circum-nuclear dust and show little evidence of ongoing starformation \citep{bal08}. A compilation of these 
properties is needed for larger samples of LEG FR-IIs to make stronger inferences. 

The HEG FR-IIs on the other hand nicely form an FR-II population that are the unbeamed counterparts of the broad-line
radio galaxies and quasars \citep{chi02}. The strong ionizing continuum and presence of gas and dust in these sources 
result in the strong narrow (low and high excitation) lines and broad lines where as the absence or only weak presence 
of a strong continuum and gas and dust 
in the FR-Is and LEG FR-IIs prevents emission of strong narrow lines and high excitation lines. The absence of
a broad line region in FR-Is has also been linked to the higher gas temperatures \citep{bot10} where dense gas clouds 
may not form. 

We may point to the interesting issue of prevalence of FR-I quasars (\citet{hey07} and reference therein). 
Having been identified as quasars the hosts must have high ionization as well as
broad emission line spectra. This implies that the accretion rates are high and the disks are radiatively efficient. It is
possible that (as also speculated by \citet{hey07}) the resulting high power beams generated may be encountering high 
enough resistance in the form of 
dense gas that makes the powerful beams lossy (as in FR-Is) and not be subject to beaming effects nor 
create edge-brightened structures where they impinge on the ambient medium.

\section{Summary}

In the work presented here we have argued for a framework within which to 
understand the two basic morphological classes, the FR-I and FR-II type radio galaxies. 
We have made a connection between host galaxy 
properties and the FR classes that has crucially
included observations of dust and the relative orientations of dust, host major axis and radio axis. 
This contributes to a picture of FR classification that now encompasses a broader range of phenomena.
With the backdrop of the Owen-Ledlow diagram we have attempted to connect various observations into 
a coherrent picture for understanding the FR dichotomy. 



Noting the quite different dust properties and relative orientations between dust, radio and major axes
of the two FR-classes as well as differences in many 
other respects and also considering their relationship with the host galaxy (as displayed in the Owen-Ledlow diagram)
we have tried to explore the physical conditions that could give rise to the seen differences. 
We are able to provide a qualitative framework for understanding the many characteristics of FR-I and FR-II radio
galaxies by considering that although massive elliptical galaxies may
generate sufficient fuel from internal sources to 
power the AGN given their high threshold jet energy flux they would more
commonly host FR-I type source morphologies and also that for lower mass ellipticals on the other hand 
the lower threshold jet energy fluxes may more easily be breached and FR-IIs more
easily generated as a result of any additional external sources of fuel (coming in through mergers or interactions).


(a) The frequently seen alignments between dust, major axis and radio axis in FR-Is and their frequent association with
more massive ellipticals suggest stable fueling conditions without need for mergers. The alignments suggest  
that they may have been left relatively unperturbed for a long enough time to have their black holes realign with the 
minor axes. However, for the FR-IIs, such undisturbed conditions may have eluded them.

(b) Mergers appear crucial for the formation of FR-IIs where as FR-Is form in 
more benign conditions.

(c) The observation of mostly aligned representations of multiple activity epochs in restarted radio
galaxies suggests that while mergers could have perturbed black hole axes in FR-II hosts the cause
for the interruption and re-triggering of the AGN is less perturbing and may not be due to a new merger event. 
The aligned structures demand that the black hole axis be steady within a merger event. 
A possible cause has been identified in the form of an assocation of
the interruption and re-triggering to the fueling as each tranche of fuel is exhausted. 
Moreover, the timescale for the entire beam activity (including the quiescent phase and renewed activity) 
is likely small compared to the realignment timescale.

(d) LEG FR-IIs, a class that shares with the FR-Is a state of the AGN characterized by a weak central 
ionizing source may include sources that have waned in AGN activity. At least some LEG FR-IIs are the 
transition FR-IIs (whether dying or restarting). A bigger compilation of LEG FR-IIs properties is needed 
to investigate their nature.

In the framework the association of an FR-type with an elliptical galaxy is
flexible depending on the power of the jets that are created in the prevailing 
conditions, the type of environment they encounter as well as the age of the activity; mass and history 
of the host elliptical galaxy play a key role.
A combination of host galaxy mass, its environment, 
the merger history, dust acquisition and distribution,  
accretion rates, ambient environment of the jets, black hole realignments may all need to be considered in
understanding the FR-I and FR-II characteristics that we observe.
We have tried to elucidate how these could influence the AGN and the radio source 
it generates. 

Radio galaxies with powers at the extremes and hosted by ellipticals of a fixed absolute magnitude will
be interesting to study to look at factors responsible for the different radio powers and morphologies.

\acknowledgments

\clearpage


\begin{thebibliography}{}
\bibitem[Baldi \& Capetti(2008)]{bal08} Baldi, R. D., Capetti, A., 2008, A\&A, 489, 989
\bibitem[Baum et al.(1992)]{bau92} Baum, S. A., Heckman, T. M., van Breugal, W., 1986, \apj, 389, 208
\bibitem[Baum et al.(1995)]{bau95} Baum, S. A., Zirbel, E. L., O'Dea, C. P., 1995, \apj, 451, 88
\bibitem[Best \& Longair(2005)]{bes05} Best, P. N., Kauffmann, G., Heckman, T. M., Brinchmann, J., 
Charlot, S., Ivezic, Z., \& White, S. D. M., 2005, \mnras, 362, 25
\bibitem[Best \& Heckman(2012)]{bes12} Best, P. N., Heckman, T. M., 2012, \mnras, tmp.2402B
\bibitem[Bicknell (1986)]{bic86} Bicknell, G. V., 1986, \apj, 305, 109
\bibitem[Bicknell (1995)]{bic95} Bicknell, G. V., 1995, \apjs, 101, 29
\bibitem[Blandford \& Rees(1974)]{bla74} Blandford, R.D., Rees, M.J., 1974, \mnras, 169, 395
\bibitem[Blundell et al.(1999)]{blu99} Blundell, K. M., Rawlings, R., Willott, C. J., 1999, \aj, 117, 677
\bibitem[Browne \& Battye(2011)]{bro11} Browne I. W. A., Battye, R. A., 2010, in "Accretion and Ejection 
in AGN: a Global Viewa", eds.  L. Maraschi, G. Ghisellini, R. Della Ceca, and F. Tavecchio, 
ASPC, 427, 365
\bibitem[Buttiglione et al.(2010)]{bot10} Buttiglione, S., Capetti, A., Celotti, A., Axon, D. J., Chiaberge, M., 
Macchetto, F. D., Sparks, W. B., 2010, A\&A, 509, 6
\bibitem[Caproni et al.(2006)]{cap06} Caproni, A., Abraham, Z., Mosquera Cuesta, H. J., 2006, \apj, 638, 120
\bibitem[Chiaberge et al.(1999)]{chi99} Chiaberge, M., Capetti, A., \& Celotti, A., 1999, \mnras, 349, 77
\bibitem[Chiaberge et al.(2000)]{chi00} Chiaberge, M., Capetti, A., \& Celotti, A., 2000, A\&A, 355, 873
\bibitem[Chiaberge et al.(2002)]{chi02} Chiaberge, M., Capetti, A., \& Celotti, A., 2002, A\&A, 394, 791
\bibitem[de Koff et al.(2000)]{dek00} de Koff, S. et al. 2000, \apjs, 129, 33
\bibitem[de Ruiter et al.(2002)]{deR02} de Ruiter, H. R., Parma, P., Capetti, A., Fanti, R., \& Morganti, R., 2002,
A\&A, 396, 857
\bibitem[Di Matteo et al.(2003)]{dim03} Di Matteo, T., Allen, S.W., Fabian, A.C., Wilson, A.S., Young, A.J., 2003,
\apj, 582, 133
\bibitem[Emsellem et al.(2007)]{ems07} Emsellem, E., et al. 2007, \mnras, 379, 401 
\bibitem[Falle (1991)]{fal91} Falle, S.A.E.G., 1991, \mnras, 250, 581
\bibitem[Fanaroff \& Riley(1974)]{fan74} Fanaroff, B. L., \& Riley, J. M., 1974, \mnras, 167, 31
\bibitem[Ghisellini \& Celotti(2001)]{ghi01} Ghisel lini, G., Celotti, A., 2001, A\&A, 379, L1
\bibitem[Gopal-Krishna \& Wiita(2000)]{gop00} Gopal-Krishna, \& Wiita, P. J., 2000, A\&A, 363, 507
\bibitem[Govoni et al. (2000)]{gov00} Govoni, F., Falomo, R., Fasano, G., Scarpa, R., 2000, A\&A 353, 507
\bibitem[Guthrie (1980)]{gut80} Guthrie, B., N., G., 1980, \apss, 70, 211
\bibitem[Hardcastle et al.(2007)]{har07} Hardcastle, M. J., Evans, D. A., Croston, J. H., 2007, \mnras, 376, 1849
\bibitem[Heckman et al.(1986)]{hec86} Heckman, T. M. et al. 1986, \apj, 311, 526
\bibitem[Heywood et al.(2007)]{hey07} Heywood, I., Blundell, K. M., Rawlings, S., 2007, \mnras, 381, 1093
\bibitem[Hill \& Lilly(1991)]{hil91} Hill, G. J., \& Lilly, S. J., 1991, \apj, 367, 1
\bibitem[Hine \& Longair(1979)]{hin79} Hine, R.G., Longair, M. S., 1979, \mnras, 188, 111
\bibitem[Ho (2009)]{ho09} Ho, L., 2009, \apj, 699, 626
\bibitem[Hopkins et al. (2011)] {hop11} Hopkins, P. F., Hernquist, L., Hayward, C. C., Narayanan, D., 2011,
\mnras, arXiv1111.1236H
\bibitem[Kaiser et al.(1997)]{kai97} Kaiser, C. R., Dennett-Thorpe, J., Alexander, P., \mnras, 292, 723
\bibitem[Kaiser \& Best (2007)]{kai07} Kaiser, C.R. \& Best, P.N., 2007, \mnras, 381, 1548
\bibitem[Kawakatu et al. (2009)]{kaw09} Kawakatu, N., Kino, M., Nagain, H., 2009, \apjl, 697, 173
\bibitem[Kormendy (2009)]{kor09} Kormendy, J., Fisher, D. B., Cornell, M. E., Bender, R., 2009 \apjs, 182, 216
\bibitem[Kotanyi \& Ekers(1979)]{kot79} Kotanyi, C. G., \& Ekers, R. D., 1979, \mnras, 73, L1
\bibitem[Laing et al.(1994)]{lai94} Laing, R. A., Jenkins, C. R., Wall, J. V., Unger, S. W., 1994, 
in "The First Stromlo Symposium: The Physics of Active Galaxies.", eds.  G.V. Bicknell, M.A. Dopita, and P.J. Quinn,
ASP Conference Series, 54, 201
\bibitem[Ledlow \& Owen (1996)]{led96} Ledlow, M. J., \& Owen, F. N., 1996, \aj, 112, 9
\bibitem[Ledlow (1997)]{led97} Ledlow, M. J., 1997, in "The Nature of Elliptical Galaxies; 2nd 
Stromlo Symposium ", eds. G. S. Da Costa; and P. Saha, ASPC, 116, 421
\bibitem[Lin et al. (2010)] {lin10} Lin Y-T., Shen, Y., Strauss, M. A., Richards, G. T., Lunnan, 
R., 2010, \apj, 723, 1119
\bibitem[Mahony et al. (2011)]{mah11} Mahony, E., K. et al. 2011, \mnras, 417, 2651
\bibitem[Marchesini et al.(2004)]{mar04} Marchesini, D., Celotti, A., \& Ferrarese, L., 2004, \mnras, 351, 733
\bibitem[McNamara and Nulsen (2007)]{mcn07} McNamara, B.R., Nulsen, P.E.J., 2007, \araa, 45, 117
\bibitem[Natarajan \& Pringle(1998)]{nat98} Natarajan, P., \& Pringle, J. E., 1998, \apjl, 506, 87
\bibitem[Owen \& Laing(1989)]{owe89} Owen, F. N., \& Laing, R. A., 1989, \mnras, 238, 357
\bibitem[Owen \& Ledlow (1994)]{owe94} Owen, F., N., \& Ledlow, M., J. 1994, in "The First Stromlo Symposium: 
The Physics of Active Galaxies.", eds. G.V. Bicknell, M.A. Dopita, and P.J. Quinn, ASPC, Vol. 54, p319
\bibitem[Owen(1993)]{owe93} Owen, F. N., 1993, LNP, 421, 273
\bibitem[Palimaka et al. (1979)]{pal79} Palimaka, J. J., Bridle, A. H., Fomalont, E. B., Brandie, G. W.,
1979, \apjl, 231, 7
\bibitem[Ramos-Almeida et al. (2012)] {ram12} Ramos Almeida1, C. et al. 2012, \mnras, 419, 687
\bibitem[Rawlings \& Saunders(1991)]{raw91} Rawlings, S., \& Saunders, R., 1991, Nature, 349, 138
\bibitem[Sansom et al. (1987)]{san87} Sansom, A. E., Danziger, I. J., Ekers, R. D., Fosbury, R. A. E., Goss, W. M.,
Monk, A. S., Shaver, P. A., Sparks, W. B., Wall, J. V., 1987, \mnras, 229, 15
\bibitem[Saripalli \& Subrahmanyan(2009)]{sar09} Saripalli, L., \& Subrahmanyan, R., 2009, \apj, 695, 156
\bibitem[Saripalli et al.(2012)]{sar12} Saripalli, L., Subrahmanyan, R., Thorat, K., Ekers, R. D., Hunstead, R. W., 
Johnston, H. M., Sadler, E. M., 2012, \apjs, 199, 27
\bibitem[Scarpa \& Urry(2001)]{sca01} Scarpa, R., \& Urry, M. C., 2001, \apj, 556, 749
\bibitem[Scheuer (1974)]{sch74} Scheuer, P. A. G., 1974, \mnras, 176, 513
\bibitem[Smith \& Heckman(1989)]{smi89} Smith, E. P., \& Heckman, T. M., 1989, \apj, 341, 658
\bibitem[Tadhunter et al.(1998)]{tad98} Tadhunter, C. N., Morganti, R., Robinson, A., Dickson, R., Villar-Martin, M.,
Fosbury, R. A. E., 1998, \mnras, 298, 1035
\bibitem[van~Dokkum \& Franx(1995)]{van95} van Dokkum, \& P. G., Franx, M., 1995, \aj, 110, 2027
\bibitem[Verdoes Kleijn \& de Zeeuw (2005)]{ver05} Verdoes Kleijn, G. A., de Zeeuw, P. T., 2005, A\&A, 435, 43
\bibitem[Wold et al.(2007)]{wol07} Wold, M., Lacy, M., \& Armus, L., 2007, A\&A, 430, 531

\end{thebibliography}
\end{document}